\documentclass{PoS}

\usepackage{amsmath}
\usepackage{bm}
\usepackage{braket}
\usepackage{fancyref}
\usepackage{cite}
\usepackage{booktabs} 
\usepackage{array} 
\usepackage{paralist} 
\usepackage{verbatim} 
\usepackage{subfig} 

\usepackage{graphicx}

\DeclareMathAlphabet{\mathcal}{OMS}{cmsy}{m}{n}

\title{On the Cosmological Frame Problem}

\ShortTitle{On the Cosmological Frame Problem}

\author{Sotirios Karamitsos
\\
        School of Physics and Astronomy, University of Manchester, Manchester M13 9PL, UK\\
        E-mail: \email{apostolos.pilaftsis@manchester.ac.uk}}

 \author{\speaker{Apostolos Pilaftsis} \\
        School of Physics and Astronomy, University of Manchester, Manchester M13 9PL, UK\\
        E-mail: \email{sotirios.karamitsos@manchester.ac.uk}}

\abstract{We introduce a fully-frame covariant formalism for inflation by taking into account conformal transformations in addition to field reparametrizations. We begin by providing a brief overview of frame problems in the history of science before outlining the crux of the frame problem in inflationary cosmology. After introducing the concept of frame tensors in curved field space, we demonstrate how the quantum perturbations and the observables sourced by them can be made frame covariant. We then specialize to two-field models, examining the impact of isocurvature effects on the inflationary observables in a frame-covariant manner. We study the phenomenology of two particular models, a minimal polynomial model and a nonminimal model inspired by Higgs inflation. We observe that in the latter scenario, isocurvature effects are greatly enhanced. Moving beyond the tree-level approximation, we outline how our approach may be extended at the quantum level through the Vilkovisky--De Witt formalism and the generalization of frame tensors to configuration space, leading to a fully frame-invariant effective action. Finally, we summarize our findings and present possible future directions of research on the topic of frame covariance.}

\FullConference{Corfu Summer Institute 2017: ``School and Workshops on Elementary Particle Physics and Gravity,'' \\ 2-28 September 2017, Corfu, Greece}

\begin{document}

\section{Brief Historical Background}

The question of whether a particular frame of reference is inherently better suited for describing the world dates back to antiquity~\cite{Waerden}. One of the first instances of this issue was born out of the clash between the \emph{geocentric} and \emph{heliocentric} models of the solar system. Anaximander
from~6~century~(c)~BC  and other pre-Socratic philosophers suggested that the Earth sits stationary at the center of the universe, while the planets, the Sun, and the stars revolve around it. Ptolemy later in 2c AD refined the ideas of geocentrism by positing that the trajectories of celestial bodies trace an \emph{epicycle} around a point which in turn orbits the Earth, a system which became the golden standard for cosmology throughout the Middle Ages and most of the Renaissance~\cite{Paschos}. In~16c~AD, however, Tycho\- Brahe's measurements of planetary motion and Galileo's observations of the phases of Venus cast doubts on the validity of the Ptolemaic model. A heliocentric model was first proposed by Aristarchus in 3c BC and further elaborated by Seleucus in 2c BC, possibly using trigonometric methods~\cite{Waerden}. But, the heliocentric  model, which was criticised by Aristotle and others, went largely un\-noticed until 15c~AD when it was geometrically formulated by Copernicus. In~16c~AD, it was further developed by Kepler in form of his titular laws of planetary motion. Even though Brahe attempted to reconcile the two models in his so-called \emph{Tychonic model}, it was not until 17c~AD that Newton's law of universal attraction provided a solid theoretical underpinning for Kepler's laws. Then, geocentric models were finally eclipsed and the heliocentric system saw widespread acceptance.

The history of the geocentric and heliocentric systems was characterized by the gradual realization of the idea that the Earth does not occupy a special place in the Universe, often referred to as the \emph{Copernicean principle}. We may view this principle as a statement of the fact that \emph{no observer is more privileged than any other}, an idea which was crucial in the development of Einstein's theory of relativity. We may glimpse this notion as early as in the Tychonic model, in which the Earth is orbited by the Sun which is in turn orbited by Mercury and Venus, as well as the rest of the planets in their own orbits. This model is related to our modern picture of the solar system by means of a coordinate transformation, and as such, the two are entirely equivalent in the sense that it is impossible to distinguish between them via observations. While certain frames might be more ``physically appealing'' (for instance, the non-inertial Earth frame of reference in the Tychonic model necessarily requires the 
consideration of {\em fictitious forces}), there is no reason to prefer one system over the other apart from convenience and ease of calculation.

\begin{figure}[t]
  \centering
\hspace{-1em} 
\includegraphics[scale=0.5]{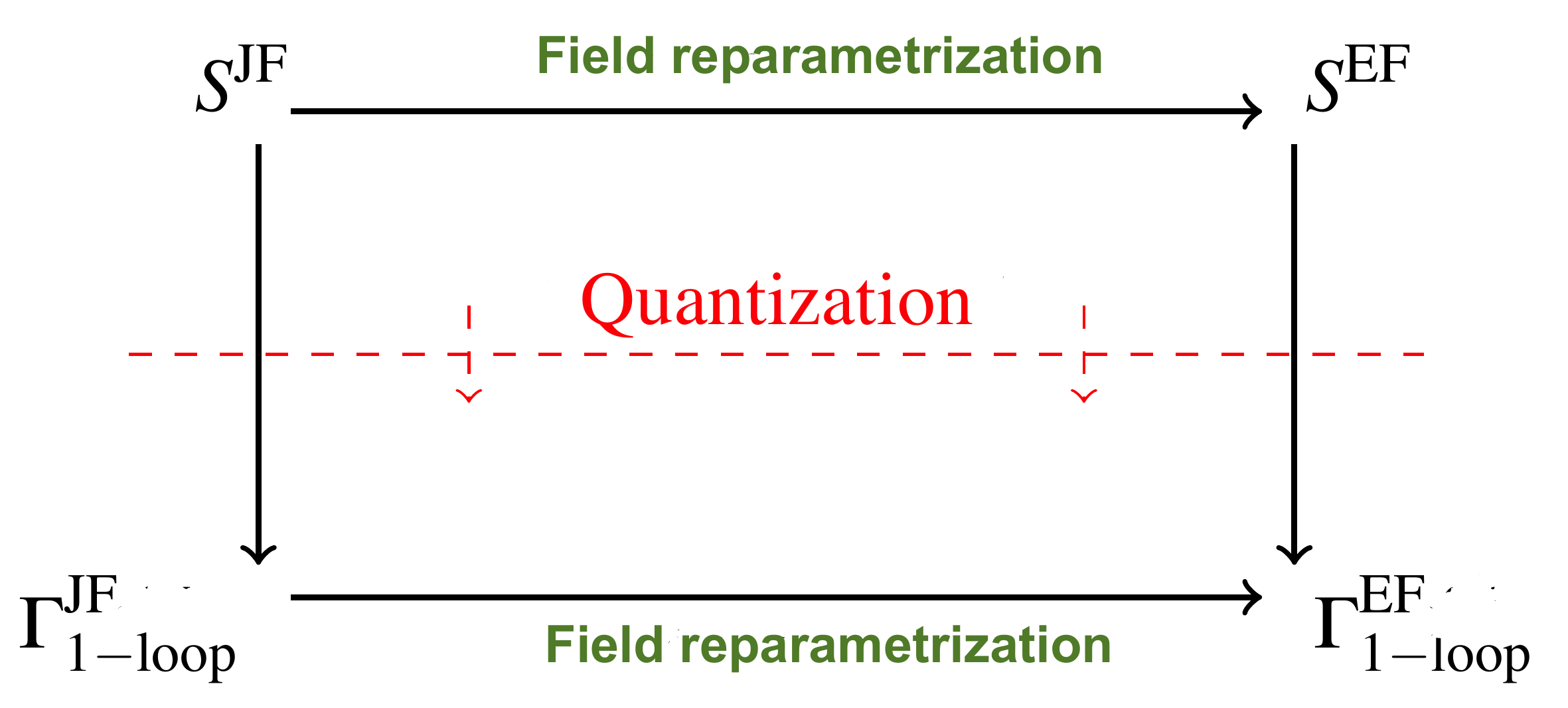} 
\caption{Illustration of quantization in different frames.  The conventional effective action formalism leads to $\Gamma^{\rm JF}_{\rm 1-loop}[g_{\mu\nu}, \varphi] \neq \Gamma^{\rm EF}_{\rm 1-loop}[\widetilde g_{\mu\nu}, { \widetilde \varphi}]$, even though $S^{\rm JF}_{\rm 1-loop} [g_{\mu\nu}, \varphi] = S^{\rm EF}_{\rm 1-loop}[\widetilde g_{\mu\nu}, { \widetilde \varphi}]$.}
  \label{fig:framequant}
\end{figure}

In this report, we extend the notion of frame equivalence to frame \emph{covariance} in the context of scalar-curvature theories with the aim of resolving the frame problem in inflationary cosmology. The action of such theories can be written in the \emph{Jordan} frame~$S^{\rm JF}[g_{\mu\nu},\varphi]$ or the \emph{Einstein} frame~$S^{\rm EF}[g_{\mu\nu},\varphi]$ depending on whether the minimal coupling $f(\varphi)$ is made explicit: 
\begin{align} 
\begin{aligned} S^{\rm JF}[ g_{\mu\nu}, \varphi] \; &= \; \int_x \left[ -\frac{1}{2} f( \varphi) R \; + \; \frac{1}{2}(\nabla_\mu \varphi)^2 \; - \; V( \varphi)\right] \; ,
    \\
    S^{\rm EF}[\widetilde g_{\mu\nu}, { \widetilde \varphi}] \; &= \; \int_x \left[ -\frac{1}{2} M_P^2 \, \widetilde R \; + \; \frac{1}{2}(\nabla_\mu \widetilde \varphi)^2 \; - \; \widetilde V(\widetilde \varphi)\right] \; .  \end{aligned} 
\end{align} 
A \emph{frame transformation} makes it possible to go from a theory with a field-dependent effective Planck mass squared $f( \varphi)$ to one with a constant $M_P^2$. Dicke~\cite{dicke} recognized that this procedure is essentially a unit transformation. In the same way that no observer enjoys an inherent privilege, there should also be no preferable system of units, and hence we may conclude that different frames should be physically indistinguishable. However, this is not always made manifest, especially when the units are field-dependent (as reflected in the dependence of $f(\varphi)$ on the fields). In particular, when we compute the conventional one-particle irreducible effective action $\Gamma$ beyond the tree-level approximation in different frames (e.g.~Einstein versus Jordan frame), this generically leads to two different effective actions, as schematically illustrated in Figure \ref{fig:framequant}. This issue has been extensively discussed in the recent literature  \cite{Flanagan:2004bz,Chiba:2013mha,Postma:2014vaa,Burns:2016ric,Jarv:2016sow,Steinwachs:2013tr,Kamenshchik:2014waa,Domenech:2015qoa,Karam:2017zno}. Whilst it is commonly accepted that different frames are classically equivalent \cite{Flanagan:2004bz,Chiba:2013mha,Postma:2014vaa,Burns:2016ric,Jarv:2016sow}, there is still no consensus how this property can be made manifest or whether it persists after quantization~\cite{Steinwachs:2013tr,Kamenshchik:2014waa,Domenech:2015qoa,Karam:2017zno}.  We call this problem {\em the cosmological frame problem}~\cite{Burns:2016ric,Karamitsos:2017elm}.  

In order to address the cosmological frame problem, we closely follow~\cite{Karamitsos:2017elm} and introduce the concept of \emph{frame covariance} in curved field space in Section \ref{cov}, incorporating conformal transformations into the well-known reparametrization covariant formalism. In Section~\ref{fico}, we examine how quantum perturbations can be made covariant, leading to manifestly invariant cosmological observables. We move on to study two-field models in Section \ref{twofield}, taking a look at the effects of curved field space on the isocurvature modes and specializing to a minimal polynomial model and a nonminimal model inspired by Higgs inflation. Finally, we illustrate the utility of this covariant formalism by straightforwardly extending it beyond the tree level through the Vilkovisky--De Witt formalism in Section~\ref{beyondtree} before outlining our conclusions and potential directions for future research in Section~\ref{conclusions}.

\section{Frame Covariance in Curved Field Space}
  \label{cov}
Standard Big-Bang cosmology is plagued by a number of problems, the most prominent of which are the near-flatness of our Universe and the largeness of the causal horizon, both of which require a high degree of fine tuning of the initial conditions of the Universe. A period of accelerated\- expansion in the early Universe not only resolves these issues in a natural way, but also provides a quantitative framework for explaining the origin of anisotropies in the CMB. A particularly interesting\- class of inflationary models is scalar-curvaure multifield inflation. In scalar-curvature models, we assume that only scalar fields are light enough to drive inflation and that they are not necessarily minimally coupled to the Ricci scalar. Such theories are described by the following classical action written in the Jordan frame:
\begin{align}
   \label{actionJ}
S  \equiv    \int  d^4 x\,  \sqrt{-g}  \, \left[  - \frac{f(\varphi)}{2} R + \frac{k_{AB}(\varphi)}{2} \, g^{\mu\nu }(\nabla_\mu \varphi^A) (\nabla_\nu \varphi^B)  -  V(\varphi)   \right] \;.
\end{align}
In this notation, uppercase indices~$A, B, \ldots$  run over the different fields, $\varphi$ without indices collectively stands for all the inflaton fields when appearing as an argument,~$g_{\mu\nu}$ is the metric function, $g\equiv \det g_{\mu\nu}$, and $R$ is the Ricci scalar. The model functions $f(\varphi), k_{AB}(\varphi)$, and $V(\varphi)$ can be selected to specialize to a wide array of models. It is possible to rewrite the action by applying a \emph{frame transformation}, which consist of a \emph{conformal transformation}:
\begin{align}
 \label{conftrans}
  \begin{alignedat}{3}
&g_{\mu\nu} \ &&\rightarrow \  \tilde g_{\mu\nu}  \ &&=\  \Omega^2  g_{\mu\nu} \; ,\\
&\varphi^A \ &&\rightarrow \ \widetilde\varphi^A  \ &&=\  \Omega^{-1} \varphi^A \; ,
\end{alignedat}
\end{align}
followed by a \emph{field reparametrization}:
\begin{align}
 \label{fieldrepam}
  \begin{alignedat}{3}
&\varphi^A &&\rightarrow \varphi^{\tilde A}  \ &&=\ \varphi^{\tilde A} ({\bf \varphi}) \; ,
\\
&\frac{d\widetilde\varphi^{\widetilde A}}{d\varphi^B} &&\rightarrow \varphi^{\tilde A}  \ &&=\ \Omega^{-1} K^{\widetilde A}_{\ B} \; .
\end{alignedat}
\end{align}
Under a frame transformation \eqref{conftrans} and \eqref{fieldrepam}, the model functions transform as
\begin{equation}
\label{transrules}
\begin{aligned}
\tilde f(\widetilde \varphi) \  &=\ \Omega^{-2}\,f(  \varphi) \;, \\ 
{\tilde k}_{ \widetilde A  \widetilde B}(\widetilde\varphi)\ &=\   \left[k_{AB}(\varphi)   - 6   f \Omega^{-2 }  \Omega_{,A} \Omega_{,B} 
  + 3 \Omega^{-1 }   f_{,A}\Omega_{,B} 
  +3\Omega^{-1} \Omega_{,A}  f_{,B} \right] K^A_{\  \widetilde A} \, K^B_{\ \widetilde   B}\;,
  \\
 \widetilde V(\widetilde \varphi)\  &=\ \Omega^{-4}\,V(\varphi)\;.
\end{aligned}
\end{equation}
These transformation rules may be used to show that the action~$S$ is form invariant, i.e.
\begin{align}
S[\tilde g_{\mu\nu}\,,\, \widetilde \varphi\,,\, \tilde f (\widetilde \varphi)\,,\, \tilde k_{ \widetilde A  \widetilde B}(\widetilde \varphi)\,,\, \widetilde V(\widetilde \varphi)] \  =\   S[g_{\mu\nu}\,,\, \varphi\,,\, f(\varphi)\,,\, 
k_{AB} (\varphi)\,,\,V(\varphi)]  \; .
\end{align}
Thus, theories related by a frame transformation define an equivalence class, which is our starting point for introducing the idea of frame covariance. 

Turning our attention to $k_{AB}$, we observe that, although it does not transform covariantly, we may use it to introduce a tensorial quantity which does, the \emph{field space metric} $G_{AB}$:
\begin{align}
   \label{eq:GAB}
G_{AB}\  &\equiv\ \frac{k_{AB}}{f} + \frac{3}{2} \frac{f_{,A} f_{,B}}{f^2}\; .
\end{align}
This quantity transforms as follows under a frame transformation:
\begin{align}
\widetilde G_{\widetilde A \widetilde B}\  &= \Omega^{2} \ G_{AB}\, K^A_{\ \widetilde A}K^B_{\ \widetilde B}  \; .
\end{align}
The above transformation motivates us to define \emph{frame tensors} in field space by their transformation properties under a frame transformation:
\begin{align}
\label{covdef}
\widetilde  X^{\widetilde A_1 \widetilde A_2 \ldots \widetilde A_p}_{\widetilde B_1 \widetilde B_2 \ldots \widetilde B_q} \ &=\ \Omega^{-(w_X + p-q)}  (K^{\widetilde A_1}_{\ A_1} K^{\widetilde A_2}_{\ A_2} \ldots K^{\widetilde A_p}_{\ A_p} )  
\ X^{  A_1   A_2 \ldots A_p}_{B_1 B_2 \ldots B_q} \ 
(K^{ B_1}_{\ \widetilde B_1}K^{ B_2} _{\ \widetilde B_2}\ldots K^{ B_q}_{\ \widetilde B_q})\;,
\end{align}
where the \emph{conformal weight} $w_X$ of a quantity $X$ describes how the quantity scales under conformal transformations, whereas the \emph{scaling dimension} describes how it scales under frame transformations, including field reparametrizations. Values of $w_X$ and $d_X$ are given in Table~\ref{tab:confweight} for various covariant quantities. Using this definition of a frame tensor given in \eqref{covdef}, we may define fully frame-covariant differentiation with respect to the fields as follows:
\begin{equation}
\begin{aligned}
\label{fieldcovderdef}
\nabla_C X^{A_1 A_2\ldots A_p}_{B_1 B_2\ldots  B_q }\ \equiv\  X^{A_1 A_2\ldots A_p}_{B_1 B_2\ldots B_q,C} & - \frac{w_X}{2} \frac{f_{,C}}{f} X^{A_1 A_2\ldots A_p}_{B_1 B_2\ldots B_q} \\ 
&+ \ \Gamma^{A_1}_{CD} X^{D A_2\ldots A_p}_{B_1 B_2\ldots B_q}   
\ +\  \cdots \, +   \Gamma^{A_p}_{CD} X^{A_1 A_2\ldots D}_{B_1 B_2\ldots B_q}
\\
& - \ \Gamma^{D}_{B_1 C} X^{A_1 A_2\ldots A_p}_{D B_2\ldots B_q} 
- \cdots
- \Gamma^{D}_{B_q C} X^{A_1 A_2\ldots A_p}_{B_1 B_2 \ldots B_q} \;.
\end{aligned}
\end{equation}
This form for the derivative respects both field reparametrizations (thanks to the terms in the last two lines) and conformal transformations (thanks to the second term in the first line). We may further define frame-covariant differentiation with respect to an arbitrary parameter $\lambda$ as follows:
\begin{align}
\label{paramcovder}
\mathcal{D}_\lambda X^{A_1 A_2 \ldots A_p}_{B_1 B_2\ldots B_q}\ 
&\equiv\  \frac{d \varphi^C}{d \lambda} \,
\nabla_C X^{A_1 A_2 \ldots A_p}_{B_1 B_2\ldots B_q}\; .
\end{align}
Using the frame-covariant forms of derivatives given in \eqref{fieldcovderdef} and \eqref{paramcovder}, we may write down the scalar field equations of motion and the Friedmann equation under the assumption of a FRW metric given by $g_{\mu\nu} = {\rm diag}(N_L^2, -a^2, -a^2,-a^2)$, where the covariant Hubble parameter is $\mathcal{H} = (\mathcal{D}_t a)/a$ and $U \equiv V/f^2$:
 \begin{align}
 \label{invarinfleq}
&  \mathcal{D}_t \mathcal{D}_t{\varphi}^A
+    3\mathcal{H}     (\mathcal{D}_t {  \varphi}^A )
+   f   G^{AB} U_{,B}\ =\ 0\;,
\\
\mathcal{H}^2\ &=\   \frac{1}{3 } \left( \frac{G_{AB} (\mathcal{D}_t \varphi^A )(\mathcal{D}_t \varphi^B)}{2 } +  f U \right).
 \end{align}
We may also write the frame-invariant form of the \emph{number of e-folds}:
 \begin{align}
N \equiv \int_{t_{\rm end} }^t dt' \, \mathcal{H}(t') \;.
 \end{align}
The equations of motion are manifestly frame-covariant, and are crucial in determining the evolution of the perturbations that form the seeds for the observable cosmological anisotropies, which will be the topic of the next section.
 
\begin{table}
\centering
\begin{tabular}{ lcc   }
$X$	 			& conformal weight ($w_X$)& scaling dimension ($d_X$)			 \\	
\hline
$dx^\mu$			&   $\hphantom{-}0$ 	&  $\hphantom{-}0$		 		\\
$d\varphi^A$		&   $\hphantom{-}0$ 	&  $\hphantom{-}1$				\\
$d\varphi_A$		&   $\hphantom{-}0$ 	&  $-1$						\\
$g_{\mu\nu}$		&   $-2$ 			&  $-2$	 					\\
$g^{\mu\nu}$		&   $\hphantom{-}2$ 	&  $\hphantom{-}2$				\\
$N_L,a$			&   $-1$ 			&  $-1$	 					\\
$  \mathcal{H} = (\mathcal{D}_t a)/a$		&   $\hphantom{-}1$	&  $\hphantom{-}1$		 		\\
$f$				&   $\hphantom{-}2$    	&  $\hphantom{-}2$		 		\\ 
$G_{AB}$  			&   $\hphantom{-}0$	&  $-2$		 				\\
$G^{AB}$  			&   $\hphantom{-}0$	&  $\hphantom{-}2$				\\
$U \equiv V/f^2$  				&  $\hphantom{-} 0	$	&  $\hphantom{-}0$		 		\\
$X^{A_1 A_2 \ldots A_p}_{B_1 B_2 \ldots B_q}$ 			& $w_X $			&  $w_X +p-q $	\\
$\nabla_AX^{A_1 A_2 \ldots A_p}_{B_1 B_2 \ldots B_q}$ 			& $w_X $			&  $w_X-1+p-q $	\\
$\mathcal{D}_\lambda X^{A_1 A_2 \ldots A_p}_{B_1 B_2 \ldots B_q}$	&$w_X-d_{\delta \lambda}$		&$w_X-d_{\delta \lambda} +p-q$ \\
\end{tabular}
 \caption{Conformal weights and scaling dimensions of various frame-covariant quantities.}
  \label{tab:confweight}
\end{table}

\section{Frame Invariant Cosmological Observables}
\label{fico}

The cosmological perturbations that eventually go on to source the profile of the Cosmic Microwave Background~(CMB) are seeded by the correlation functions of the primordial perturbations of the metric and the scalar fields. Metric perturbations can be parametrized to first order in the scalar-vector-tensor decomposition in the Newtonian gauge as follows:
\begin{align}
\label{metricpar}
g_{\mu\nu} dx^\mu dx^\nu\ =\ (1+2\Psi)N_L^2 \, dt^2\: -\: a^2 \big[ (1-2\Phi)\delta_{ij} + h_{ij}\big] dx^i dx^j.
\end{align}
 We may further define the frame-covariant extensions of the gauge-invariant \emph{Mukhanov-Sasaki} variables~\cite{Sasaki:1986hm,Mukhanov:1988jd}:
\begin{align}
\label{sasakivar}
Q^A\ \equiv\ \delta \varphi^A + \frac{\mathcal{D}_t \varphi^A}{\mathcal{H}} \Phi\; .
\end{align}
The concept of the \emph{field space} arises naturally when looking at the trajectories which satisfy the equations of motion \eqref{invarinfleq} and when viewing the fields $\varphi^A$ as \emph{coordinates} parametrizing a manifold~\cite{Gong:2011uw, Elliston:2012ab}. The simplest choice for the metric is given by \eqref{eq:GAB}, which also naturally leads to the definition of the field-space connection $\Gamma^A_{BC}$ and the line element~$d\sigma^2 = G_{AB}\,d\varphi^A d\varphi^B$. The field space formalism is useful in the presence of isocurvature perturbations. We can decompose the perturbations $Q^A$ along the \emph{curvature} (parallel) and \emph{isocurvature} (perpendicular) directions using the vielbein fields:
\begin{align}
\label{framefields}
e_\sigma^A\  &=\ \frac{\mathcal{D}_t \varphi^A}{ \mathcal{D}_t \sigma}\; ,
&
e_{s_1}^A \ &=\ - \frac{(s_1)^{AB} \, U_{,B}}{ \sqrt{(s_1)^{AB} \, U_{,A} U_{,B} }} \; = \; -\frac{\omega^A}{\omega} \;.
\end{align} 
We focus on the ``first'' isocurvature mode $Q^{s_1}$ as it is the only mode which couples to the curvature mode. The \emph{acceleration vector} is denoted by $\omega^A$, and the projection operator to the isocurvature subspace is given by
\begin{align}
 {(s_1)}^{A B}\ \equiv\ G^{AB} - e_{\sigma}^A  e_{\sigma}^B\;.
\end{align}
In order to make contact with observations, we turn our attention to the \emph{comoving curvature perturbation} $\mathcal{R}$ and the \emph{comoving isocurvature perturbations} $\mathcal{S}^{(i)}$:
\begin{align}
\label{curvpert}
\mathcal{R}\ \equiv\ \frac{\mathcal{H}}{\mathcal{D}_t \sigma} Q^\sigma, \qquad
\mathcal{S}^{(i)}\ \equiv\ \frac{\mathcal{H}}{\mathcal{D}_t \sigma} Q^{s_i}.
\end{align}
Both $\mathcal{R}$ and $\mathcal{S}^{(i)}$ are gauge- and frame-invariant. The curvature perturbation $\mathcal{R}$ is of particular interest to us, as it remains constant on superhorizon scales and sources the (dimensionless) observable scalar spectrum $P_\mathcal{R}$ through its two-point function:
 \begin{align}
 \label{scalpowspec}
\frac{2\pi^2}{p^3} P_\mathcal{R} \, \delta^{(3)}({\bf p} + {\bf q})\ \equiv\ \braket{\mathcal{R}_{\bf p}|\mathcal{R}_{\bf q}} \; ,
\end{align}
where $\mathcal{R}_{\bf p}$ is the Fourier transform of the perturbation.
Solving and quantizing the perturbed equations of motion \cite{Sasaki:1995aw}, we find the following expressions for the power spectra under the (covariant) slow-roll approximation $\mathcal{D}_t \mathcal{D}_t {\varphi^A}  \ll \mathcal{H} (\mathcal{D}_t  \varphi^A )$:
 \begin{align}
 \label{scal}
 P_{\mathcal{R} } \ =\ \frac{ \mathcal{H}^2}{8\pi^2 f(\varphi) \bar \epsilon_H}\;,
 \qquad
 P_T\ =\ \frac{2  }{ \pi^2}  \frac{\mathcal{H}^2}{f(\varphi)}\ ,
\end{align}
in terms of the frame-invariant \emph{Hubble slow-roll parameter} $\bar \epsilon_H \; \equiv \; - \mathcal{D}_N \mathcal{H}/\mathcal{H}$. In the multifield case, isocurvature modes evolve outside the horizon and because they are coupled to the curvature modes, the power spectrum \eqref{scal} evaluated at horizon exit differs from the observable spectrum at horizon re-entry \cite{Wands:2002bn}. To first order, we may neglect the corrections to the power spectrum that are of the order of the slow-roll parameters \cite{Byrnes:2006fr} and write 
\begin{align}
\label{PR1st}
P_{\mathcal{R}}(t)\ &=\ \Big[1  +  T^2_\mathcal{RS}(t_*,t)\Big] P_{\mathcal{R}  }(t_*)\;,
\end{align}
where $ T^2_\mathcal{RS}(t_*,t)$ is the \emph{entropy transfer function}, $t_*$ is the time at horizon exit and $t$ is the time at observation.

We may now write down frame-invariant expressions for the standard inflationary observables in terms of the power spectra $P_\mathcal{R}$ and $P_T$. These are the \emph{scalar spectral index} $n_\mathcal{R} $, the \emph{tensor spectral index}~$n_T $, and \emph{tensor-to-scalar ratio} $r$:
 \begin{align}
 \label{odef1}
n_\mathcal{R} - 1\equiv\left.\frac{d \ln  P_\mathcal{R} }{d\ln k}\right\vert_{k =  a \mathcal{H} } ,
\qquad
n_T \equiv\left.\frac{d \ln  P_T }{d\ln k}\right\vert_{k =  a \mathcal{H}},
\qquad
r \equiv     \frac{  P_T }{  P_\mathcal{R} },
\end{align}
where we evaluate every parameter at the time of horizon exit $k=a \mathcal{H}$ in all expressions for the observables.
We also define the \emph{runnings} of the spectral indices as follows:
\begin{align}
\label{odef2}
\alpha_\mathcal{R} &\equiv \left.\frac{d    n_\mathcal{R} }{d\ln k}\right\vert_{k =    a \mathcal{H}}\; ,
\qquad
\alpha_T   \equiv \left.\frac{d    n_T }{d\ln k}\right\vert_{k =    a \mathcal{H}}\;,
\end{align}
as well as the \emph{non-linearity parameter}~$f_{NL}$~\cite{Komatsu:2001rj}, given by:
 \begin{align}
 \label{fNL}
f_{NL}\ =\ \frac{5}{6} \frac{  N^{,A} N^{,B} (\nabla_A \nabla_B N)}{(  N_{,A} N^{,A})^2}\;,
\end{align}
where the frame-covariant derivative $\mathcal{D}_N$ with respect to the number~$N$ of $e$-folds is defined with the help of \eqref{paramcovder}. We wish to write explicit expressions for these observables in terms of the \emph{potential slow-roll parameters}, defined through the following hierarchy:
\begin{equation}
\label{xisrp}
\begin{aligned}
\bar\epsilon_{U,1} \ &\equiv \ \bar\epsilon_U\;,
\\
&\ \ \vdots
\\
\bar \epsilon_{U,n} \  &\equiv\ -\frac{(\bar \epsilon_{U,n-1})_{,A}}{\bar \epsilon_{U,n-1}}  G^{AB} \frac{U_{,B}}{U}\ ,
\end{aligned}
\end{equation}
where $\bar\epsilon_{U,1} \equiv \bar\epsilon_U$, $\bar\epsilon_{U,2} \equiv \bar\eta_U$ and  $\bar\epsilon_{U,3} \equiv \bar\xi_U$. Thus, the cosmological observables in the slow-roll approximation become
\begin{equation}
\begin{aligned} 
n_\mathcal{R} &=  1 - 2 \bar \epsilon_U +  \bar \eta_U - \mathcal{D}_N \ln \big(1+ T_\mathcal{RS}^2\big)\;,
&
n_T &=  -2\bar\epsilon_U\;,
&
r &= 16 \bar \epsilon_U (\cos\Theta)^2\;,
\\
\alpha_\mathcal{R}  &=     
 - 2 {\bar \epsilon}_U  {\bar\eta}_U  - {\bar \eta}_U {\bar \xi}_U +  \mathcal{D}_N \mathcal{D}_N \ln \big(1+ T_\mathcal{RS}^2\big)\;,
&
\alpha_T &= 
-2\bar\epsilon_U\bar\eta_U\; .
\end{aligned}
\end{equation}
In a similar vein, we may obtain a simple formula for $f_{NL}$  by substituting the following expression for $N_{,A}$ in \eqref{fNL}:
 \begin{align}
N_{,A}\ =\ \frac{U\, U_{,A}}{U_{,B}\, U^{,B}}\;.
\end{align}
We have thus written down concise expressions for the cosmological observables. We now specialize to models with only two scalar degrees of freedom, examining the effects of entropy transfer on the observables before looking at particular examples of two-field models.

\section{Two-Field Models}
\label{twofield}

Turning our attention to the entropy transfer between the modes, we find that solving the superhorizon equations of motion is in general very difficult. We may, however, derive approximately analytic results for two-field models. In this case, the isocurvature modes are fully encoded in~$\mathcal{S} \equiv \mathcal{S}^{(1)}$, and the superhorizon equations of motion simplify to~\cite{Lalak:2007vi,Kaiser:2012ak,vandeBruck:2016rfv}:
\begin{equation}
\label{superhorizontwofield}
\begin{aligned}
\mathcal{D}_N \mathcal{R}\ &=\  -2\omega\, \mathcal{S}\;, \\
\mathcal{D}_N \mathcal{S}\ &=\ - B(N)\,   \mathcal{S}\; ,
\end{aligned}
\end{equation}
where the \emph{turn rate} $\omega$ (which is the norm of the acceleration vector $\omega^A$) controls the rate of isocurvature transfer and the model-dependent parameter $B(N)$ controls the generation of isocurvature modes. To lowest slow-roll order, the acceleration vector is
\begin{align}
  \label{turnrateapprox}
 \omega^A\ &=\  (\ln U)^{,B}\ \nabla_B \left[\frac{(\ln U)^{,A}}{\sqrt{2 \bar\epsilon_U}}\right].
\end{align}
In the constant slow-roll approximation where $\omega = \omega_*$ and $B=B_*$ are evaluated at horizon crossing, the transfer function $T_{\mathcal{RS}}$ becomes
\begin{align}
\label{TRSold}
 T_{\mathcal{RS}}(N_*, N)\ &\approx \ \frac{2\omega_* }{B_* } \Big[e^{ -B_*  (N-N_*) } - 1 \Big]\ .
\end{align}

\begin{figure}[t]
  \centering
  \hspace{-1em}
\includegraphics[scale=0.9]{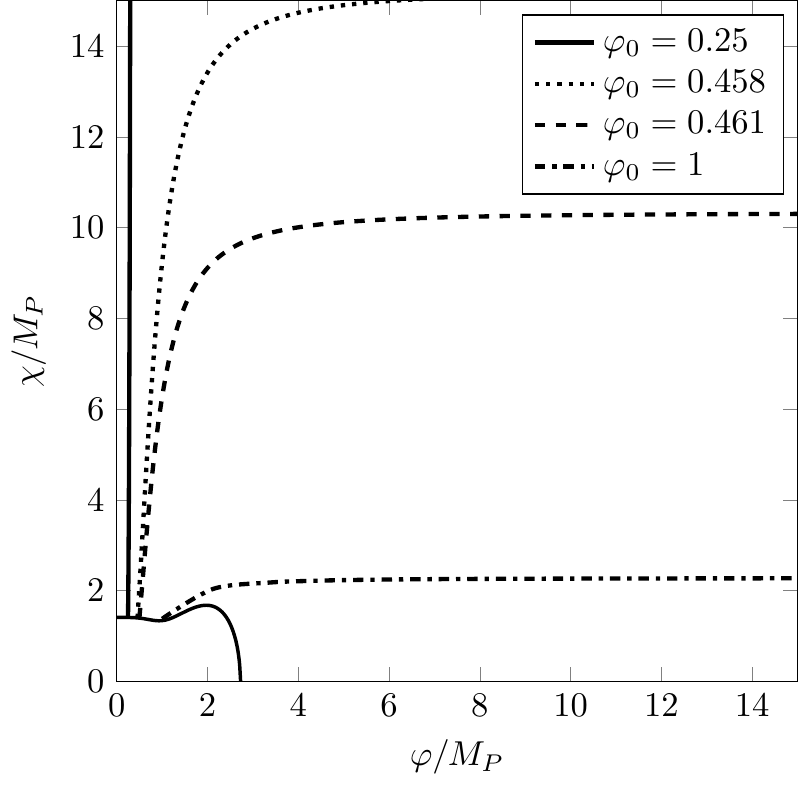}
 \includegraphics[scale=0.9]{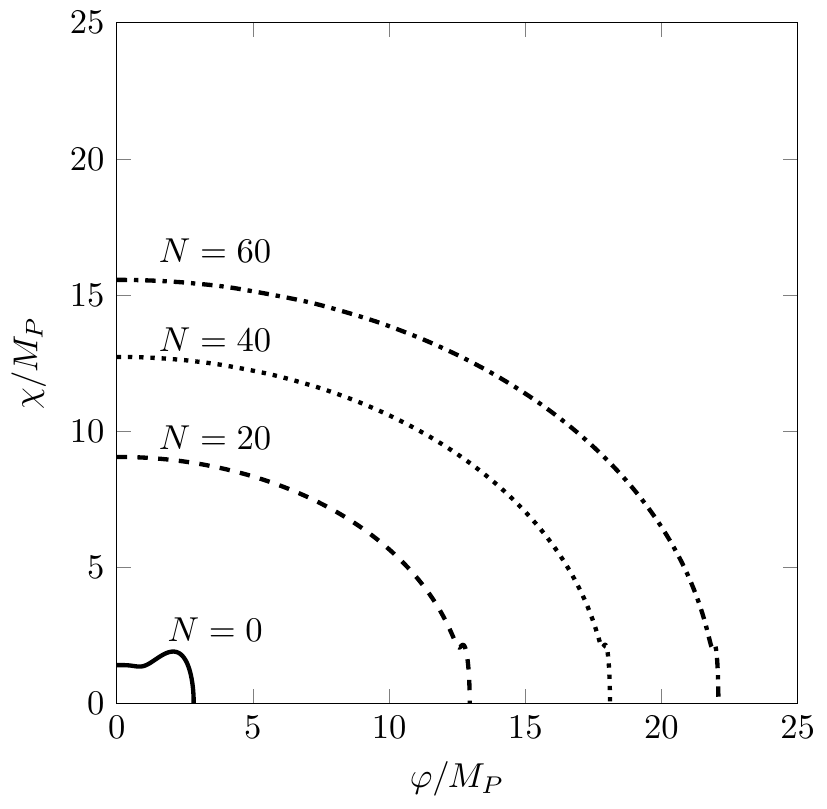}
  \caption{Field space trajectories and isochrone curves for the minimal model.}
  \label{fig:trajectories}
\end{figure}

\begin{figure}[t]
  \centering
  \hspace{-1em}
\includegraphics{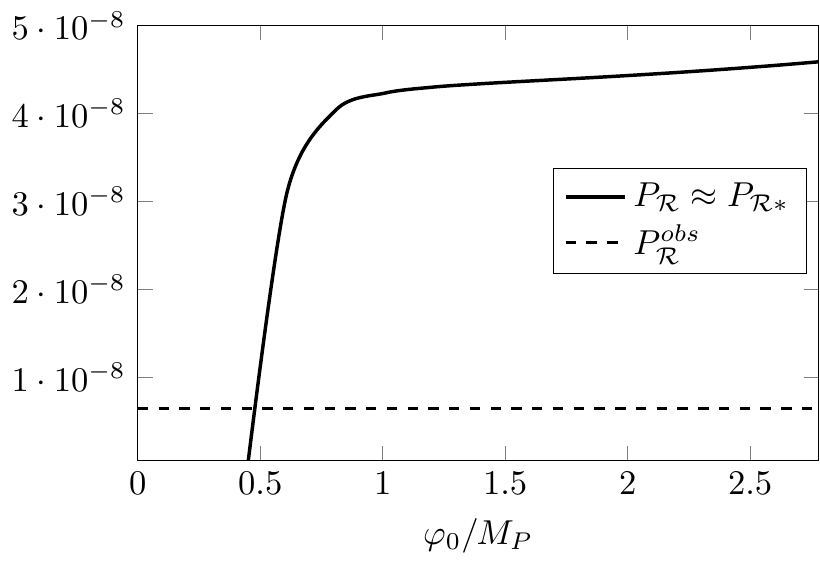}
  \caption{Power spectrum normalization for the minimal model with $\lambda  = 10^{-12}$ and $m/M_P = 10^{-6}$ at~$N=60$ for different boundary conditions in terms of $\varphi_0$.}
  \label{fig:PRMIN}
\end{figure}

\begin{figure}[t]
  \centering
\includegraphics{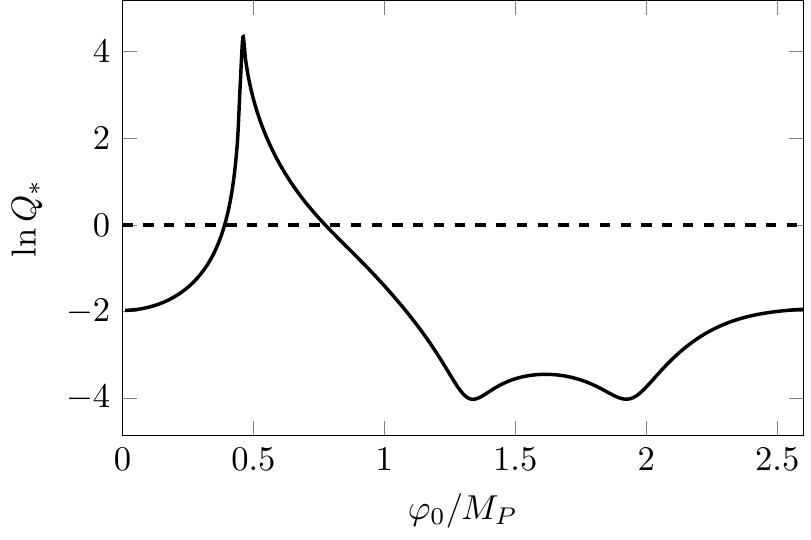}
  \caption{Sensitivity parameter $Q_*$  for the minimal model at $N=60$ to boundary conditions given by $\varphi_0$. }
  \label{fig:qMIN}
\end{figure}

We may apply the above results to two particular models, starting with a minimal two-field model described by the following action:
\begin{align}
\label{lagr1}
\mathcal{L} = -\frac{ M_P^2 R}{2} + \frac{1}{2} (\nabla \varphi)^2 +  \frac{1}{2} (\nabla \chi)^2 - \frac{\lambda\varphi^4 }{4}  - \frac{m^2 \chi^2}{2} \;,
\end{align}
where $m$ is a mass parameter and $\lambda$ is a quartic coupling~\cite{GarciaBellido:1995qq}. Solving the equations of motion for different boundary conditions, the field trajectories change as shown in Figure~\ref{fig:trajectories}.
By matching the value of $P_\mathcal{R}$ to the observed scalar power spectrum $P^\text{obs}_\mathcal{R} = (6.41 \pm 0.18) \times 10^{-9}$ at the $68 \%$ confidence level \cite{Ade:2015lrj}, we may single out a particular observationally viable trajectory as seen in Figure~\ref{fig:PRMIN}.
We further define the \emph{sensitivity} parameter $Q_*$ of the trajectory at $N_*$ with respect to the boundary conditions at $N_0$ as the ratio of the density of trajectories on the two isochrones as follows:
\begin{align}
\label{relsens}
Q_*\   \equiv\ \frac{ 
\sqrt{\left\vert\det  [\Gamma_{IJ}]_{N_*}\right\vert}/\int_{N=  N*}   \, \sqrt{\left\vert\det  [\Gamma_{IJ}]_{N_*}\right\vert}
}{  
\sqrt{\left\vert\det  [\Gamma_{IJ}]_0\right\vert}/\int_{N=0}  \, \sqrt{\left\vert\det  [\Gamma_{IJ}]_0\right\vert}
}   \ ,
\end{align}
where the \emph{induced metric} is given by $[\Gamma_{IJ}]_N \ =\  G_{AB}  \; \left( \nabla_I \varphi^A_{N}\right) \;\left( \nabla_J \varphi^B_{N}\right)$ and $\varphi^A = \varphi^A_N$ is the parametrization of the isochrone curve at $N$ e-folds. In Figure~\ref{fig:qMIN} we plot $\ln Q_*$ as a function of $\varphi_0$, and we summarize the predictions of this model in Table \ref{tab:observablesmin} and compare our predictions to the currently observed values. 
 \begin{table} 
\centering
\begin{tabular}{ l  l  l  l}
  &   $\hphantom{-}\varphi_0/M_P= 0.495$		& \hspace{0.5em} PLANCK 2015 \\	
  \hline
  $r$ 				& $\hphantom{-}0.501 $ 					& \hspace{0.5em}$\le 0.12 \ (  95 \%   \text{ CL}) $ \\ 
  $n_\mathcal{R}$		& $\hphantom{-}0.906 $ 					& $\hphantom{-}0.968 \pm 0.006 \ (  68 \%   \text{ CL})   $  \\ 
  $\alpha_\mathcal{R}$  	& $-0.00288 $  						& $-0.003 \pm 0.008  \ (  68 \%   \text{ CL})$     \\
  $\alpha_T$ 			& $-0.0019 $ 						& $-0.000167 \pm  0.000167 \ (  68 \%   \text{ CL})$ \\ 
  $f_{NL}$			& $\hphantom{-}0.0129$					& $\hphantom{-}0.8 \pm 5.0 \  (   68 \%   \text{ CL})$ \\
\end{tabular}
\caption{Observable inflationary quantities for the minimal two-field model at $N=60$.
The value of $\alpha_T$ is derived from the consistency relation \cite{Byrnes:2006fr} with transfer angle $\Theta = 0$. 
      \label{tab:observablesmin}}
\end{table}

\begin{figure}[t]
  \centering
  \hspace{-1em}
    \includegraphics[scale=0.9]{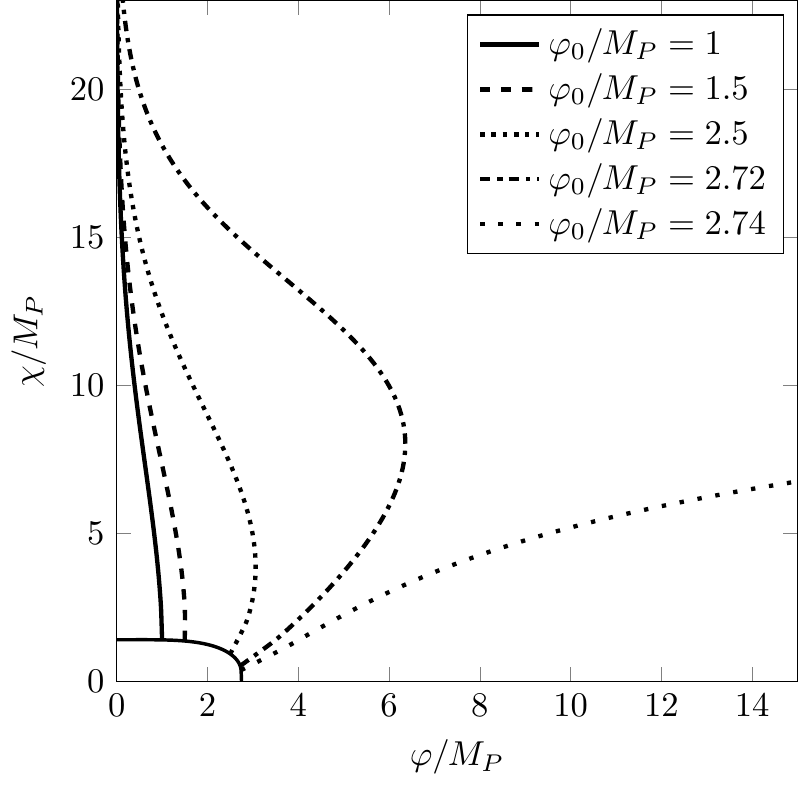}
    \includegraphics[scale=0.9]{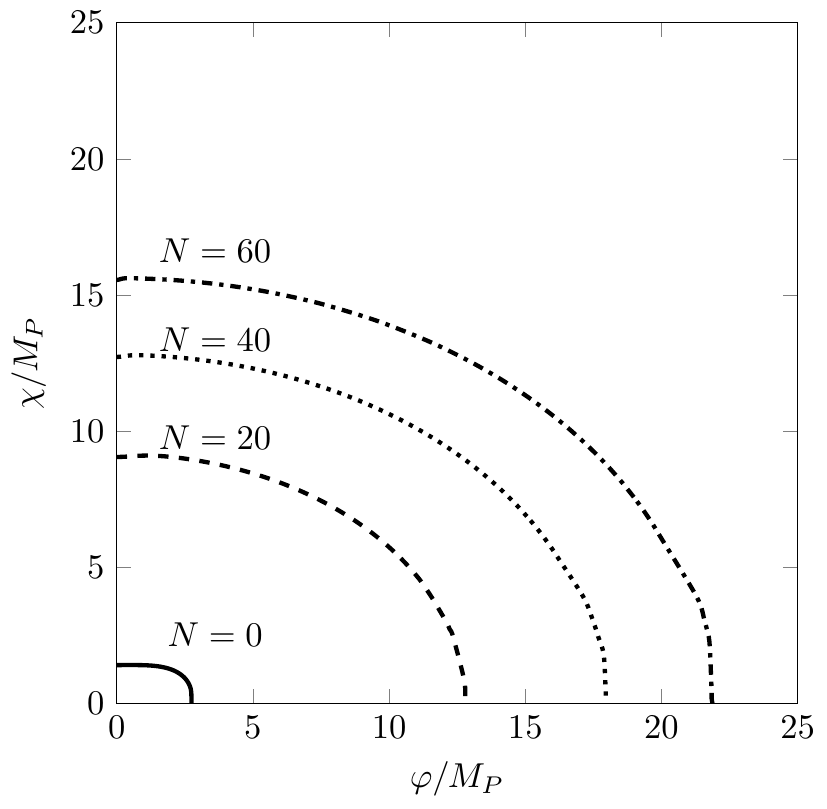}
  \caption{Field space trajectories
and isochrone curves
  for the nonminimal model.}
  \label{fig:phichiplotNM}
\end{figure}

\begin{figure} [t]
 \hspace{9.75em}
 \includegraphics{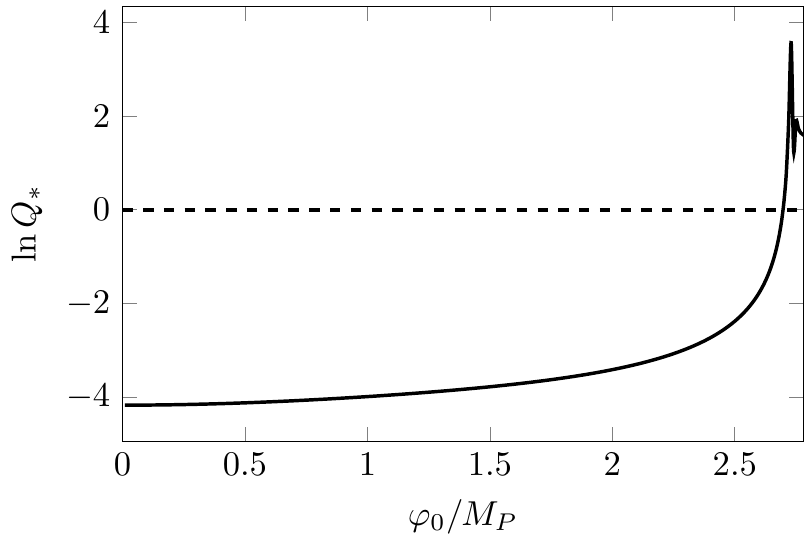}
  \caption{Sensitivity parameter $Q_*$ for the nonminimal model to boundary conditions given by $\varphi_0$.}
  \label{fig:qNM}
\end{figure}

\begin{figure}[t]
  \centering
 \includegraphics{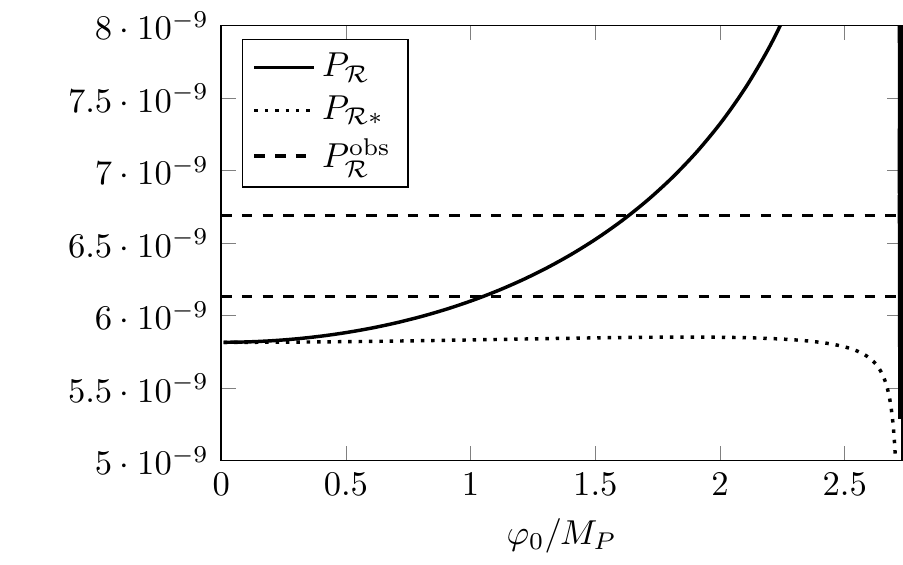}
  \caption{Power spectrum normalization for the nonminimal model with parameter values $m =5.6 \times 10^{-6} M_P$,~$\lambda = 10^{-12}$, and~$\xi = 0.01$ for different boundary conditions in terms of $\varphi_0$ and the corresponding horizon crossing values~$\varphi_*$.  }
  \label{fig:PRNM}
\end{figure}

We extend the model described by \eqref{lagr1} by including a minimal coupling $\xi$ between one of the fields and the Lagrangian, as inspired by Higgs inflation~\cite{Salopek:1988qh,Bezrukov:2007ep}.
\begin{align}
\label{lagr2}
\mathcal{L} \ = \ -\frac{(M_P^2+ \xi \varphi^2) R}{2} + \frac{1}{2} (\nabla \varphi)^2 +  \frac{1}{2} (\nabla \chi)^2 - \frac{\lambda(\varphi^2-v^2)^2 }{4}  - \frac{m^2 \chi^2}{2}\ ,
\end{align}
The field trajectories for this model are as shown in Figure~\ref{fig:phichiplotNM}, whereas the sensitivity parameter $Q_*$ can be seen in Figure~\ref{fig:qNM}. Similar to the minimal model, we plot the normalization of the nonminimal model in Figure~\ref{fig:PRNM}, observing that the inclusion of entropy transfer effects is crucial in normalizing it to observations. Finally, Table~\ref{tab:observables} shows the values of the observables for the observationally viable band of trajectories (ignoring the trivial $\varphi^4$ trajectory).

\begin{table} 
\centering
\begin{tabular}{ l  l  l  l}
               			& \quad $ \varphi_0/M_P = 1.391^{+0.243}_{-0.343}$	&\hspace{0.5em}  PLANCK 2015  \\	
\hline	
$r$ 				&  $\hphantom{-}0.1204^{+0.0053}_{-0.0049}$   		&\hspace{0.5em}$\le 0.12 \  (   95 \%   \text{ CL})$ \\ 
$n_\mathcal{R}$		&  $\hphantom{-}0.955^{+0.005}_{-0.002} $		& $\hphantom{-}0.968 \pm 0.006 \  (   68 \%   \text{ CL}) $  \\ 
$\alpha_\mathcal{R}$  	&  $-0.0004^{+0.00005}_{-0.00006}$				& $-0.008 \pm 0.008 \  (   68 \%   \text{ CL})  $    \\
$\alpha_T$ 			&  $-0.000276^{+0.000003}_{-0.000003}$			& $-0.000155 \pm  0.00016 \  (   68 \%   \text{ CL}) $    \\ 
$f_{NL}$			&  $\hphantom{-}0.0693^{+0.00003}_{-0.00002}$	 	& $\hphantom{-}0.8 \pm 5.0  \  (   68 \%   \text{ CL})  $ \\
\end{tabular}
\caption{Observable inflationary quantities for~the nonminimal model at $N=60$. 
\label{tab:observables}}
 \end{table}

\section{Beyond the Tree-Level Approximation}
\label{beyondtree}

We have demonstrated how the frame-covariant approach ensures that observables are frame independent at the classical level. With this formalism as our starting point, we wish to extend frame covariance beyond the tree level. To do so, we use the effective action formalism, beginning with the generating functional $W[J]$ written in terms of the source fields $J_a$:
\begin{align}
\exp\left\{{\frac{i}{\hbar}W[J]}\right\}\ \equiv\   \ln \int [\mathcal{D}\phi]\  \exp \left[\frac{i}{\hbar} \Big( S[\phi] +  J_a \phi^a \Big)\right]\;,
\end{align}
where the quantum field $\phi^a = \phi^A(x_A)$ lives in \emph{configuration space} and contracted indices are both summed and integrated over.
The effective action $\Gamma[\varphi]$ is then given by
\begin{align}\label{effactorig}
\exp\left(\frac{ i}{\hbar} \Gamma [ \varphi]     \right) \ =\  \int  [ \mathcal{D}\phi]\,  \exp \left\{\frac{i}{\hbar} \Big[ S[\phi] +   \Gamma_{,a} \big(\varphi^a-\phi^a\big) \Big]\right\}\,,
\end{align}
where $\Gamma_{,a} \equiv \delta \Gamma[\varphi]/\delta\varphi^a = - J_a$. The presence of the term $\big(\varphi^a-\phi^a\big)$ means that the action $\Gamma[\varphi]$ is not frame-invariant, since the fields do not transform as vectors. This may be remedied in the Vilkovisky--De Witt formalism, which replaces $\varphi^a$ by a two-point function $\Sigma^a(\varphi, \phi)$. This function is a frame-invariant scalar under a reparametrization of the quantum field $\phi$, whereas it transforms as a vector under a reparametrization of the background field $\varphi$. We may thus modify the integro-differential equation \eqref{effactorig} in order to return the Vilkovisky--De Witt effective action~\cite{Vilkovisky:1984st,Rebhan:1986wp,DeWitt}:
\begin{align}\label{effact}
\exp\left(\frac{ i}{\hbar} \Gamma [\varphi]     \right)\ =\  \int [\mathcal{D}\phi]\, \mathcal{M} [\phi]\, \exp \left\{\frac{i}{\hbar} \Big[ S[\phi]\: +\:   (\nabla_a \Gamma) \,  \Sigma^a(\varphi , \phi)   \Big]\right\} \;.
\end{align}
The measure is given by $\mathcal{M}[\varphi] \equiv \sqrt{\det \mathcal{G}_{ab}}$, and the configuration space metric (which is used to define the frame-covariant functional derivative) is given by 
\begin{align} 
\mathcal{G}_{ab} \; \equiv \; G_{AB} \delta(x_A-x_B) = \left( \frac{k_{AB}}{f} \; + \; \frac{3}{2} \frac{f_{,A}f_{,B}}{f^2}\right)  \delta(x_A-x_B) \;.
\end{align}
In order to proceed any further, we perform a perturbative $\hbar$-expansion of the frame-invariant Vilkovisky--De Witt effective action
\begin{equation}
  \label{eq:Ghbar}
\begin{aligned}
\Gamma[\varphi]\ &=\ \sum_n \hbar^n \Gamma_n\; = \; \widetilde \Gamma[\widetilde\varphi],
\end{aligned}
\end{equation}
where $\Gamma_0 [\varphi] \; = \; S[\varphi]$. This finally returns the following explicit form for the 1PI effective action:
 \begin{align}
   \label{eq:G1eff}
\Gamma_1[ \varphi ]\ &=\ -\,\frac{i}{2}\,\text{tr} \ln \mathcal{G}_{ab}\: +\:  \frac{i}{2}\, \text{tr} \ln  \Big( \nabla_a \nabla_b\, S[ \varphi] \Big)\ =\  \frac{i}{2}\, \text{tr} \ln  \Big( \nabla^a \nabla_b\, S[ \varphi] \Big)\,.
\end{align}
We may compute the corrected model parameters to higher orders in $\hbar$ by iteratively solving~\eqref{effact}. At the one-loop order, we may explicitly calculate the frame-invariant correction~$\Gamma_1[\varphi]$ to the classical action~$\Gamma_0[\varphi] = S[\varphi]$ through \eqref{eq:G1eff}.

\section{Conclusions and Future Directions}
\label{conclusions}

We have developed a fully frame-covariant formalism of inflation for multifield scalar-curvature theories at the classical level, extending it beyond the tree level so as to include radiative corrections. Making use of notions known from differential geometry, we have adopted an approach in which the scalar fields take on the role of generalized coordinates of a manifold, and the equations of motion describe a trajectory within the field space. We have studied isocurvature effects and seen that they are significant in non-minimal two-field models. We have promoted field space frame covariance to frame covariance in the configuration space of quantum fields by virtue of the Vilkovisky--De Witt formalism. Possible future research paths include applying the frame covariant approach to $F(R)$ and $F(\varphi,R)$ theories, as well as extending to include matter and quantum gravity effects. The curvature of the field space in non-minimal theories should also have an impact on cosmological observables at higher-loop orders, and these effects could have significant theoretical and phenomenological implications for theories beyond the Standard Model.

\subsubsection*{Acknowledgements}
The work of SK is supported by an STFC PhD studentship.  The work of AP is supported by the Lancaster--Manchester--Sheffield Consortium for Fundamental Physics, under STFC research grant ST/L000520/1.

\end{document}